\documentclass[aps,twocolumn,showpacs,amsmath,amssymb,pra,superscriptaddress,floatfix,longbibliography]{revtex4-1}

\usepackage{braket}
\usepackage{amsmath}
\usepackage{amssymb}
\usepackage{mathtools}
\usepackage{graphicx}
\usepackage{dcolumn}
\usepackage{bm}
\usepackage{multirow}
\usepackage{appendix}
\usepackage{leftidx}
\usepackage{color}
\usepackage{float}
\usepackage{qcircuit}
\usepackage{ifthen}
\usepackage{tabularx}
\usepackage{amsfonts}
\usepackage[german,english]{babel}
\usepackage{soul} 
\usepackage{comment}
\usepackage{afterpage}

\newcommand \be{\begin{equation}}
\newcommand \ee{\end{equation}}
\newcommand \bea{\begin{eqnarray}}
\newcommand \eea{\end{eqnarray}}
\newcommand \bse{\begin{subequations}}
\newcommand \ese{\end{subequations}}

\usepackage{xcolor}
\newboolean{ShowComments}
\setboolean{ShowComments}{false}  

\definecolor{mscolor}{rgb}{0,0.5,0.5}

\definecolor{xjcolor}{rgb}{0.5,0,0.5}

\newcommand {\rsub}[1]{\textcolor{black}{#1}}

\newcommand{\InfleqtionM}{Infleqtion, Inc., Madison, WI, 53703, USA}
\newcommand{\UWM}{Department of Physics, University of Wisconsin-Madison, 1150 University Avenue, Madison, WI, 53706, USA}

\begin{document}

\title{
Enhanced Measurement of Neutral Atom Qubits with Machine Learning}
\author{L. Phuttitarn}
\affiliation{\UWM}
\author{B. M. Becker}
\affiliation{\UWM}
\author{R. Chinnarasu}
\affiliation{\UWM}
\author{T. M. Graham }
\affiliation{\UWM}
\author{M. Saffman}
\affiliation{\UWM}
\affiliation{\InfleqtionM}

\date{\today}

\begin{abstract}
We demonstrate qubit state measurements assisted by a supervised convolutional neural network (CNN) in a neutral atom quantum processor. We present two CNN architectures for analyzing neutral atom qubit readout data: a compact 5-layer single-qubit CNN architecture and a 6-layer multi-qubit CNN architecture. We benchmark both architectures against a conventional Gaussian threshold analysis method. In a sparse  array (9 $\mu\rm  m$ atom separation) which experiences negligible crosstalk, we observed up to 32\% and 56\% error reduction for the multi-qubit and single-qubit architectures respectively, as compared to the benchmark.
 In a tightly spaced array (5 $\mu \rm m$ atom separation), which suffers from readout crosstalk, we observed up to 43\% and 32\% error reduction in the multi-qubit and single-qubit CNN architectures respectively, as compared to the benchmark. By examining the correlation between the predicted states of neighboring qubits, we found that the multi-qubit CNN architecture reduces the crosstalk correlation up to 78.5\%. This work demonstrates a proof of concept for a CNN network to be implemented as a real-time readout processing method on a neutral atom quantum computer, enabling faster readout time and improved fidelity.  
\end{abstract}
 
\maketitle

\section{Introduction}
Scalable quantum computation requires precise, high fidelity  initialization, control, and measurement of the quantum state of a large number of qubits. The number of sequential operations that can be performed is limited by qubit decoherence, resulting from imperfect qubit control mechanisms and unintended interaction with the environment. The leading approach to overcome this limitation is mid-circuit measurement based quantum error correction (see \cite{Acharya2023} and references therein).   
The ability to make high-fidelity measurements with minimal collateral disruption to the system is not only relevant to initialization and final read-out -- it is also essential to achieving quantum error correction. 

Neutral atom quantum computing has matured remarkably in recent years. Single-  and two-qubit gates have been demonstrated on neutral atoms with fidelity well above 99\% \cite{Nikolov2023,Evered2023}. Multi-qubit quantum circuits \cite{Graham2022, Bluvstein2022} and mid-circuit measurements have also been demonstrated \cite{Deist2022, Singh2023,Graham2023b,Norcia2023,SMa2023,Lis2023}.
Qubit state measurements in a neutral atom array are achieved by probing the array with light detuned from a cycling transition. The resulting fluorescence is captured with a high quantum efficiency imaging device such as an EMCCD or sCMOS sensor, producing a greyscale image of the neutral atom array. Conventionally, the state of the qubit is then determined by integrating the photon counts over regions of interest (ROIs) and applying a linear threshold that optimally separates the two states' probability distributions (see Fig. \ref{Setup}). The fidelity of the state detection is dependent upon the separability of the two distributions, which in turn depends on the signal-to-noise ratio. To achieve a fidelity above 99\% with this method, the typical probing period is tens of $\rm ms$. This is a significant delay, given that the longest gate operation only takes several $\mu \rm s$. Shortening the exposure time decreases the probing period, but also reduces signal-to-noise ratio and the fidelity. One could increase the power of the probing laser to compensate, but too much power can cause atom loss through heating. This compromise limits the signal-to-noise ratio and imposes a lower bound on the probing period, given a desired measurement fidelity and acceptable rate of qubit loss.

It is possible to further reduce the probing period without loss of fidelity by using more efficient image analysis algorithms. Examples include Independent Component Analysis\cite{Xia2015}, the Bayesian Inference Algorithm \cite{Martinez-Dorantes2017} and supervised Deep Neural Networks (DNNs)\cite{Syberfeldt2020}. In this work, we present enhanced state detection using a Convolutional Neural Network (CNN). CNNs are a sub-category of DNNs that are especially well suited for image classification. Supervised DNNs have already been demonstrated to improve readout fidelity on other leading quantum computing platforms, including 
trapped-ion \cite{Seif2018,ZHDing2019}, superconducting\cite{Bravyi2021,Lienhard2022} and quantum dot qubits\cite{Darulova2021,Matsumoto2021}. To the authors' knowledge, this is the first demonstration of the use of deep learning for neutral atom qubit state detection. 

\section{CNN Designs}

Since state detection for neutral atom qubits is performed by imaging the atoms, a CNN is ideal for the task. The principles of how CNNs work are well known and we refer to existing literature for more details \cite{Goodfellow2016}. In what follows, we focus on the particulars of the CNNs used in our experiments.

We implemented and benchmarked two different CNN architectures for qubit state detection. The first architecture, CNN-site, is a single qubit classifier. Analyzing one qubit at a time, it  replaces the conventional linear threshold method for real-time applications, achieving  superior detection fidelity. The second architecture, CNN-array, is a multi-qubit classifier. Because it analyzes the entire qubit array in parallel, it is crosstalk-aware. Compared to CNN-site, it boasts improved detection fidelity on large, tightly spaced arrays at the expense of real-time computational efficiency. Both architectures are implemented in Python using the Keras API \cite{Chollet2015} and the Tensorflow library \cite{tensorflow2015-whitepaper} with GPU acceleration.

 \begin{figure}[t]
    \includegraphics[width=\linewidth]{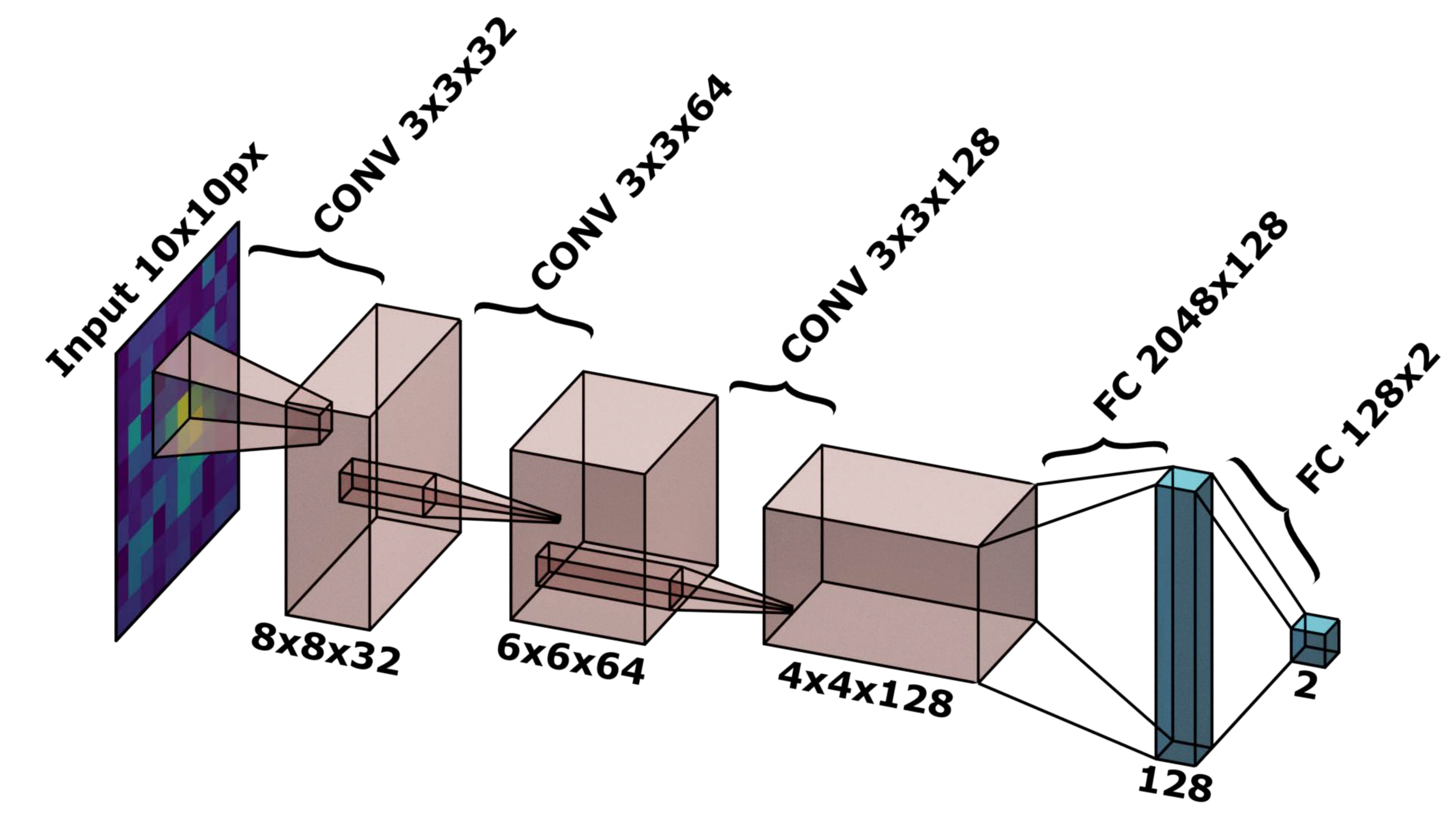}
    \caption{CNN-site network structure. The readout containing multiple sites in the array of qubits is pre-processed and sectioned into single sites before feeding into the network. The pre-processing consists of mean subtraction and normalization. For more detailed process description, see Appendix \ref{S_training}.  The single qubit readout is fed into the three convolutional layers. The final features are extracted on the first FC and connected to the output layer. The two nodes on the output layer represent the probability of the two quantum states.
   }
  \label{cnn_site}
\end{figure}
\begin{table}[t]
    \centering
    \begin{tabular}{|c|c|c|c|}
    \hline
    Layer & Filter Shape & Output Shape & \# Params\tabularnewline
    \hline 
    \hline 
    Input & N/A & $10\times10\times1$ & 0\tabularnewline
    \hline 
    Conv2D & $3\times3\times1\times32$ & $8\times8\times32$ & 320\tabularnewline
    \hline 
    Conv2D & $3\times3\times32\times64$ & $6\times6\times64$ & 18,496\tabularnewline
    \hline 
    Conv2D & $3\times3\times64\times128$ & $4\times4\times128$ & 73,856\tabularnewline
    \hline 
    Flatten & N/A & 2048 & 0\tabularnewline
    \hline 
    Dense & $2,048\times128$ & 128 & 262,272\tabularnewline
    \hline 
    Dense & $128\times2$ & 2 & 258\tabularnewline
    \hline 
    \multicolumn{3}{|r|}{Total trainable parameters:} & 355,202\tabularnewline
    \hline 
    \end{tabular}
    \label{cnn_site_parameters}
    \caption{The filter shape, output shape, and number of trainable parameters in each layer of CNN-site.}   
\end{table}

\begin{figure}[t]
    \includegraphics[width=\columnwidth]{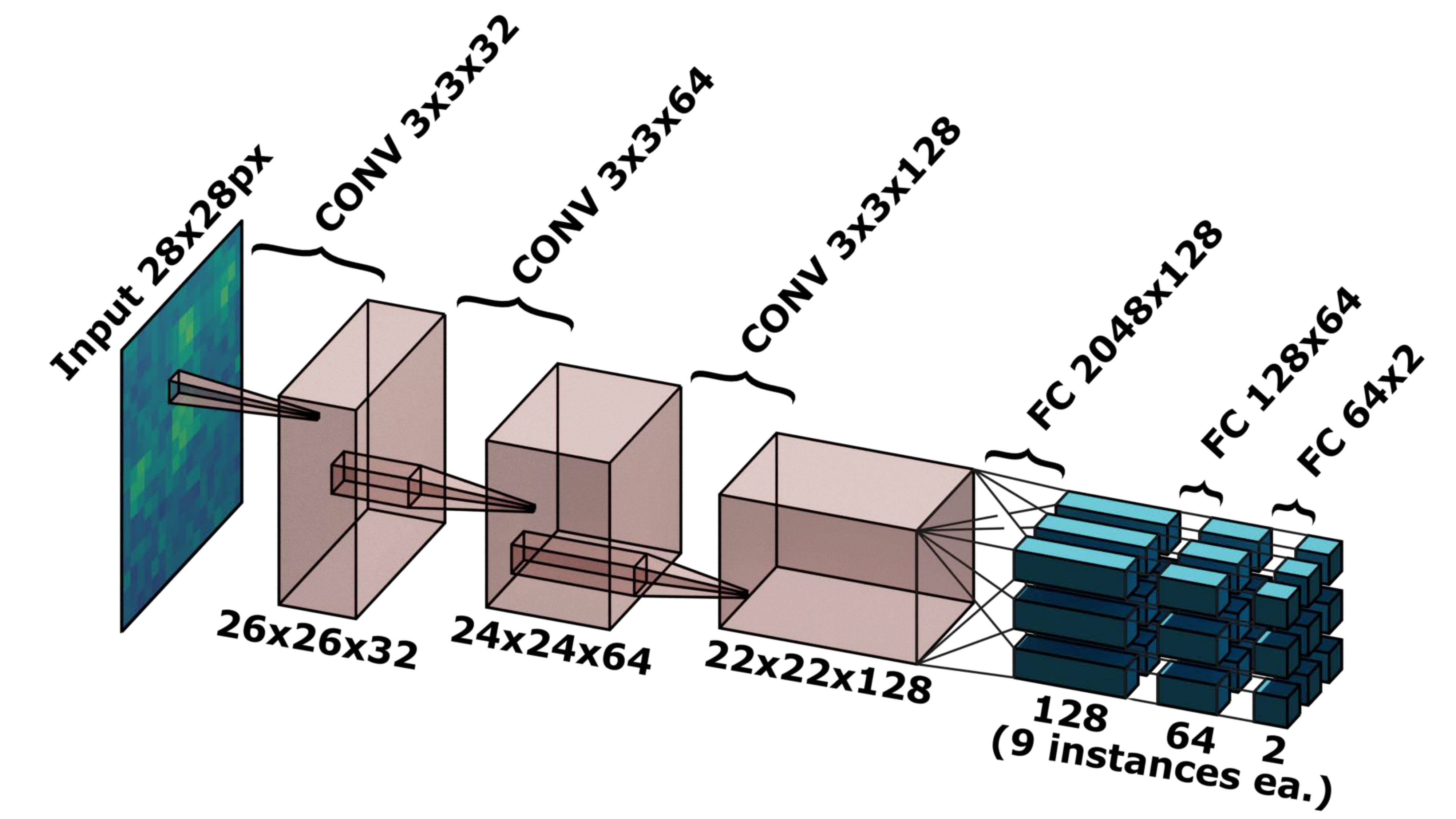} 
    \caption{CNN-array network structure. The raw readout image is preprocessed and fed directly into the convolutional layers. The features detected by the last convolutional layers are connected to multiple FC networks, each responsible for the classification of a single qubit. For the 3$\times$3 array, there are 9 independent FC networks.}
    \label{cnn_array}
\end{figure}

\begin{table}[t]
    \centering

    \begin{tabular}{|c|c|c|c|}
    \hline
    Layer & Filter Shape & Output Shape & \# Params \tabularnewline
    \hline 
    \hline 
    Input & N/A & $28\times28\times1$ & 0\tabularnewline
    \hline 
    Conv2D & $3\times3\times1\times32$ & $26\times26\times32$ & 320\tabularnewline
    \hline 
    Conv2D & $3\times3\times32\times64$ & $24\times24\times64$ & 18,496\tabularnewline
    \hline 
    Conv2D & $3\times3\times64\times128$ & $22\times22\times128$ & 73,856\tabularnewline
    \hline 
    Flatten & N/A & 61952 & 0\tabularnewline
    \hline 
    Dense & ($2048\times128)\times 9$ & $128\times 9$ & 74,613,888\tabularnewline
    \hline 
    Dense & $(128\times64) \times 9$ & $64 \times 9$ & 74,304\tabularnewline
    \hline 
    Dense & $(64\times2) \times 9$ & $2 \times 9$ & 1,170\tabularnewline
    \hline 
    \multicolumn{3}{|r|}{Total trainable parameters:} & 74,782,034\tabularnewline
    \hline 
    \end{tabular}
  \caption{The output shape and number of trainable parameters in each layer in CNN-array. Compared to the CNN-site, the number of trainable parameters is substantially higher($\sim$210 times).}
  \label{cnn_array_parameters}
\end{table}

\subsection{CNN-site}
CNN-site is a compact, five-layer single-qubit state classifier, as shown in Fig. \ref{cnn_site}. It consists of three convolutional (CONV) layers followed by two fully-connected (FC) layers. The activation function is a rectified linear unit ($g\left(\mathbf{z}\right)=\max\left(\mathbf{0},\mathbf{z}\right)$) for all layers except the final FC layer, which uses a Softmax activation function\footnote{The Softmax activation function is defined as 
$g\left(z_{c^{\prime}}\right)=\frac{e^{z_{c^{\prime}}}}{\sum_{d=1}^{C_{\rm out}}e^{z_{d}}}$
This function normalizes the sum of the elements such that $\sum_{c^{\prime}=1}^{C_{\rm out}}z_{c^{\prime}}=1$. Each element value is then the weighted probability. }. As input, it receives a cropped monochromatic image of a single site of the atomic array. The input images are pre-processed by mean subtraction and normalization (see Appendix \ref{S_training} for details). The output nodes represent the weighted probabilities of the atom being in a dark or bright state.  

\rsub{The dimensions of the CNN layers were experimentally varied in order to discover the set of parameters that consistently converged to a high state detection fidelity during training. Appreciably shrinking the network resulted in sporadic convergence failures (as indicated by low fidelities corresponding to random guessing)  due to nondeterministic initialization of the network weights. On the other hand, appreciably increasing the width or depth of the network did not yield further fidelity improvements -- the fidelities were almost identical amongst networks that did converge.} This reflects the relative simplicity of the input images; the task can be completed with only 3 convolutional layers because higher-complexity features, which additional convolutional layers could extract, are  not present in the data. \rsub{We believe that CNN-site could be further optimized without any detrimental impact to detection fidelity, but this would require further iterative fine-tuning and would not alter the conclusions of this work.} The number of parameters is already very small in comparison to typical CNNs. Its low parameter count makes it quick enough for real-time applications, including atom rearrangement and mid-circuit measurement if accelerated using a neural network accelerator (NNA) or an FPGA.

\subsection{CNN-array}
 CNN-array is a six-layer multi-qubit state classifier that improves state detection in closely-spaced arrays where crosstalk between the neighboring sites is appreciable. As input, CNN-array accepts pre-processed images of the full atomic array.  Its architecture, pictured in Fig. \ref{cnn_array}, is identical to that of CNN-site except (a) the input image is larger, (b) there is an additional FC layer, and (c) there are multiple instances of each FC layer, one per qubit in the array. It produces $\mathcal{N}$ simultaneous outputs with the same format as CNN-site, indicating the two-state probabilities of each qubit in the array. Since each instance of the FC layers can ``see'' the features of all of the other qubits in the array, CNN-array can identify and accommodate correlations between neighboring qubits. Owing to the larger input image and the all-to-all fanout of the FC layers' connections, CNN-array contains $\sim$210 times more parameters than CNN-site, making real-time implementation challenging, even with an NNA or FPGA. Optimization is likely possible and is a subject for future work.

\subsection{Gaussian-threshold \& Square-threshold} \label{conventional_methods}
For completeness, we present the conventional Gaussian threshold and square-threshold methods, which we used as our performance benchmark. 

First, the region of interest of each atom (the location in the image where the atom is expected to be found) for each site is determined by averaging all the readout images in the entire dataset, and applying a circular Gaussian fit to the result. This locates the center of the distribution, corresponding to the central position  of the atom, and the standard deviation $\sigma$, corresponding to the size of the atom in pixels.  For the Gaussian-threshold method, the resulting best-fit parameters are then used to generate a 2D-Gaussian mask for each site: a value in the range 0.0 to 1.0 for each pixel in the image. For the square-threshold method, we replace the Gaussian mask with a binary square mask with sides of length $2\sigma$. 

\begin{figure}[!t]
    \centering
    \includegraphics[width=\columnwidth]{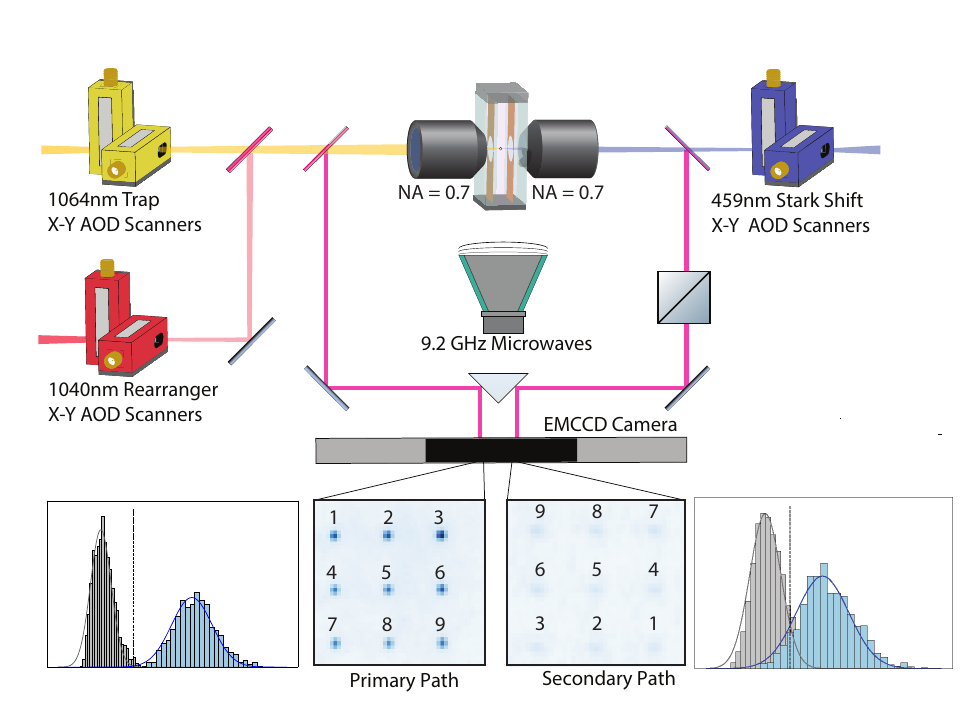}
    \caption{(Top) Experimental layout for dual path readout collection. Atoms are cooled and loaded into a $3\times 3$ array of optical tweezers. To readout the qubit state, the  array is illuminated with 852-nm light, and transition fluorescence is collected via two opposing high NA lenses. The two collection paths are projected onto separate regions of the EMCCD sensor. A non-polarizing beam splitter is placed in the secondary path to attenuate the photon collection and generate noisy data for benchmarking. The non-attenuated path is used to generate labels for supervised learning and benchmarking. (Bottom) The averaged readout image and histogram from a 10 ms integration, the shortest readout time in the dataset. The solid line represents the two Gaussian fits used in conventional analysis methods. Conventionally, a threshold(shown as dashed line) is computed that optimally separates the two Gaussian distributions, which is used to infer the qubit state. The state detection fidelity depends entirely on the separability of the two distributions. Note that the histograms pictured are based on an ensemble of images. The threshold is computed based on this ensemble, whereas state detection is performed on the camera counts from a  single image of a single atom at a time. The count distribution has a state separation of 4.5$\sigma$ and 2.1$\sigma$ from the primary and secondary paths respectively. Using the Gaussian threshold method, a state-detection fidelity $>$99.8\% is maintained in all datasets for the non-attenuated path. The combined attenuation elements result in a secondary path fluorescence efficiency 
     of 60\% of the primary path.
    }
    \label{Setup}
\end{figure}

The mask is multiplied element-wise with each readout image and then summed over all pixels, returning a single integrated value per atom. The quantum state of the atom is predicted to be a bright (dark) state if the integrated value is above (below) a pre-determined threshold. The threshold itself is determined by collating the integrated data of many images of randomly-loaded arrays. The histogram of the collated data is shown in Fig. \ref{Setup} (lower-left, lower-right) along with the threshold (dotted line). The distribution resembles a mixture of two Gaussian distributions, one for each binary state (dark or bright). The parameters of the two-Gaussian mixture can be obtained by applying a curve fit to the histogram. The threshold that maximizes the separation between the two states is the intersection point of the two fitted Gaussian curves.

\section{Experimental Methods}
To collect data, Cs atoms are loaded into a 3$\times$3 tweezer array generated by a two-dimensional Acousto-Optical Deflector (AOD). The array is imaged  with 852 nm light that is red-detuned from the $6s_{1/2}, f=4 - 6p_{3/2}, f=5$ transition by 9$\gamma$ ($\gamma = 2 \pi \times 5.2$ MHz). Fluorescence light is  collected via the two opposing 0.7 NA objective lenses. State measurements are performed destructively by first pushing out one of the states, followed by occupancy imaging with a hyperfine repump laser to prevent atoms going dark due to Raman transitions. Although state measurements can be performed non-destructively \cite{Kwon2017} the destructive method used here provides for the lowest possible uncertainty in state measurements. 
The two collection paths are projected onto two adjacent regions of the EMCCD sensor (Andor Ixon 897).

In order to train and evaluate the performance of the two CNN networks, we acquired a dataset consisting of noisy input data paired with the corresponding ``ground truth'' labels. 
We collected data at site spacings of 5 $\mu$m and 9 $\mu$m, varying the readout time(secondary path state separation) from 10 ms(2.17$\sigma$) to 100 ms(66.6$\sigma$). The conventional linear threshold method is applied to the high-SNR images in order to generate the ``ground truth'' labels with $>99.8\%$ fidelity in all sets. Each dataset contains 3,000 single-shots of the $3\times3$ array with $\sim 50\%$ occupancy for each readout.

\rsub{We do not expect the small labeling error in the dataset to create an upper bound on achievable fidelity. Convolutional neural networks and deep neural networks are known to be robust against small amounts of unbiased mislabelled data \cite{Fard2017, Rolnick2018}. The fraction of labelling errors in our dataset ($<0.2\%$) is small enough to not impose an upper limit on the achievable fidelity with a  supervised learning method. For a system with lower readout fidelity such that there is a significant amount of mislabelled data, conventional training methods can be modified with a transfer learning method \cite{Oquab2014, Azizpour2015}, semi-supervised learning \cite{Weston2012} and unsupervised learning \cite{Dosovitskiy2014}.} 

Each dataset is then randomly divided into three subsets with the distribution: 60\% training,  20\% validation, and 20\% test. The training subset is used as input for all iterations of CNN training. The validation subset is used between training epochs to monitor the performance of the partially-trained CNN and detect overfitting. The test subset is reserved for final performance evaluation, to ensure that the CNNs are benchmarked on data that the  CNNs have no prior access to. For specifics about the CNN training procedure, see Appendix \ref{S_training}. All quoted fidelities were evaluated on the test subset.

\section{Results}

  \begin{figure}[t]
    \includegraphics[width=\columnwidth]{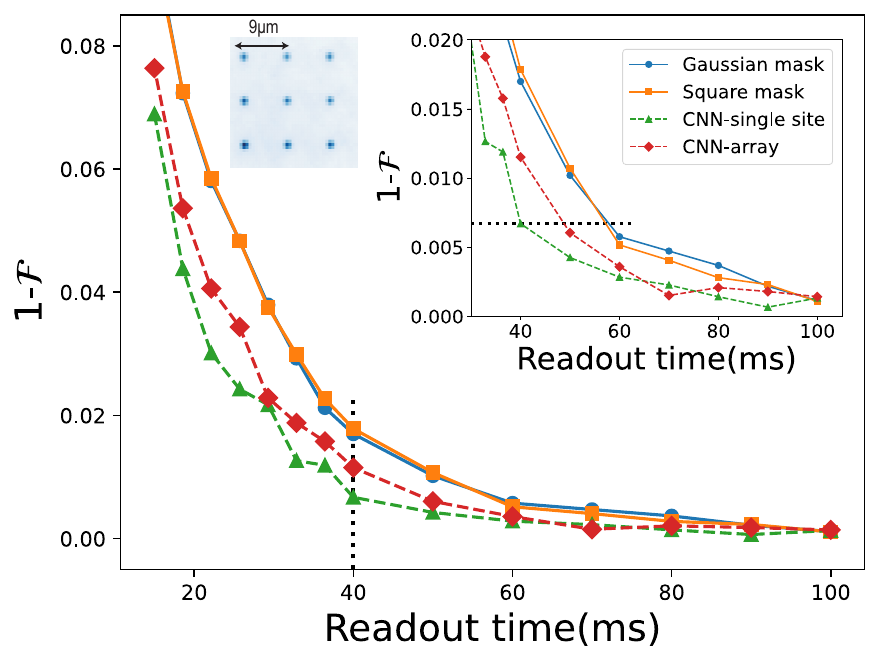}
   \caption{The measurement infidelity from different analysis methods at 9 $\mu$m site spacing. The greatest error reduction between CNN-site and Gaussian method is observed at 40 ms, marked with a vertical dashed line. The inset shows the details at readout time above 30 ms. The horizontal dashed line presents the biggest reduction of readout time on CNN-site  while maintaining the same fidelity compared to Gaussian method.}
    \label{9um_result}
\end{figure}

\begin{table}[t]
    \centering
    \begin{tabular}{c|c|c|c|c} 
        \multicolumn{5}{c}{Central to Nearest Neighbor}\tabularnewline
        \hline 
        \hline 
         & Gaussian & Square & CNN-site & CNN-array \tabularnewline
        \hline 
        \hline 
        $\mathcal{F}^{CF}_{52}$ & 0.0218 & 0.0072 & 0.0048 & 0.0043\tabularnewline
        \hline 
        $\mathcal{F}^{CF}_{54}$ & 0.0300 & 0.0267 & 0.0236 & 0.0238\tabularnewline
        \hline 
        $\mathcal{F}^{CF}_{56}$ & 0.0075 & 0.0025 & -0.0090 & -0.0048\tabularnewline
        \hline 
        $\mathcal{F}^{CF}_{58}$ & 0.0065 & 0.0186 & 0.0046 & 0.0095\tabularnewline
        \hline 
        \hline 
        \textbf{$\braket{|\mathcal{F}^{CF}_{5j}|}$} & \bf{0.0165} & \bf{0.0138} & \bf{0.0060} & \bf{0.0082}\tabularnewline
        \hline 
        \multicolumn{5}{c}{}\tabularnewline
        \multicolumn{5}{c}{Edge-to-Edge}\tabularnewline
        \hline
        \hline
         $\mathcal{F}^{CF}_{13}$ & -0.0086 & -0.0131 & -0.0059 & -0.180\tabularnewline
        \hline 
        $\mathcal{F}^{CF}_{79}$ & -0.0134 & -0.0172 & -0.0108 & -0.0118\tabularnewline
        \hline 
        $\mathcal{F}^{CF}_{17}$ & 0.0024 & 0.0151 & -0.0004 & 0.0069\tabularnewline
        \hline 
        $\mathcal{F}^{CF}_{39}$ & 0.0095 & 0.0040 & 0.0007 & 0.0004\tabularnewline
        \hline 
         $\mathcal{F}^{CF}_{19}$ & -0.0098 & -0.0108 & -0.0163 & -0.0160\tabularnewline
        \hline 
         $\mathcal{F}^{CF}_{37}$ & -0.0030 & -0.0006 & 0.0007 & -0.0055\tabularnewline
        \hline 
        \hline 
        \textbf{$\braket{|\mathcal{F}^{CF}_{ij}|}$} & \bf{0.0101} & \bf{0.0078} & \bf{0.0058} & \bf{0.0096}\tabularnewline
        \hline 
        \hline 
    \end{tabular}
    \caption{Cross-Fidelities at 40 ms readout time in the $9~\mu\rm m$ spacing array. (Top) Cross-fidelity between the central site and the nearest neighbor sites. (Bottom) Cross-fidelity between the edge sites. The cross-fidelity between central to nearest neighbors are comparable to that of edge-to-edge suggesting no observable crosstalk is observed at this separation in this configuration.}
    \label{9um-crosstalk-table}
  \end{table}

 In this section, we present the CNN classification performance compared to  conventional methods. We first consider the performance under low-crosstalk conditions (9 $\mu \rm m$ array spacing), then examine the case when there is increased crosstalk (5 $\mu \rm m$ array spacing). Crosstalk occurs when a percentage of photons emitted by an atom strike the EMCCD array at the location corresponding to a different site in the atomic array. This increases the photon count for the ``wrong'' atom, making correct state detection more challenging.
 
 Performance is measured in terms of classification fidelity \cite{Lienhard2022}, defined as
\begin{equation}
    \mathcal{F} = 1 - \frac{P(B_p|D)+P(D_p|B)}{2}
\end{equation}
where $D$ and $B$ are the true dark and bright  qubit states respectively and $D_p$ and $B_p$ are the predicted dark and bright states. $P(B_p|D)$ and $P(D_p|B)$ represent the false bright and false dark prediction probabilities respectively. To quantify the prediction correlation between two sites due to crosstalk, we define the cross-fidelity \cite{ZChen2021} as 
\begin{equation}
    \mathcal{F}^{\rm CF}_{ij} = 1-\braket{P(D_i|B_j)+P(B_i|D_j)}
\end{equation}
where $D_i$ and $B_i$ are the dark and bright states of  qubit $i$ and  $D_j$ and $B_j$ are the dark and bright states of qubit $j$. In the absence of crosstalk, the state of qubit $i$ is independent of the qubit $j$, and the probability of qubit $i$ being in the bright state given that qubit $j$ is in the dark state, $P(D_i|B_j)$, is 0.5. This yields $\mathcal{F}^{CF}_{ij}=0$ for the crosstalk-free case. Positive (negative) $\mathcal{F}^{CF}_{ij}$ values indicate correlation (anti-correlation) between sites $i$ and $j$.

We define the relative infidelity factor
$\eta$ in terms of CNN infidelity $(1-\mathcal{F}_{\rm CNN}$) and Gaussian mask infidelity $(1-\mathcal{F}_{\sigma}$) as 
\begin{equation}
\eta=\frac{(1-\mathcal{F}_{\sigma})-(1-\mathcal{F}_{\rm CNN})}{1-\mathcal{F}_{\sigma}}
\end{equation}
This measures the infidelity reduction, or the percent reduction in incorrect state predictions, that the CNN achieved as compared to the conventional method.

Figure \ref{9um_result} shows the performance of all four state detection methods for the $9~\mu \rm m$ spacing array as a function of readout time. At this spacing, the effects of crosstalk are relatively small, which we confirmed by comparing the cross-fidelity of the neighboring sites versus non-neighboring sites for the Square mask in Table \ref{9um-crosstalk-table}. (We use the Square mask statistics as a reference because that method has no built-in mechanism to mitigate crosstalk). Both CNN architectures showed a reduction in measurement infidelity relative to the conventional methods at readout times up to 90 ms. CNN-site achieved up to 56\% reduction in infidelity (vertical dotted line), or up to 29\% reduction in readout time (horizontal dotted line) when compared to the Gaussian method. CNN-array achieved up to 35\% reduction in infidelity and 14\% reduction in readout time. This demonstrates that the CNNs outperform the conventional method even in a low-crosstalk configuration. CNN-site's superior performance, despite its low parameter count, may indicate that it is better-optimized for this particular task than CNN-array. In this crosstalk-free configuration, the large number of trainable parameters in CNN-array does make it more susceptible to overfitting. We also observe no significant differences in performance between the Square Threshold and Gaussian Threshold.

 \begin{table}[t]
       \begin{tabular}{c|c|c|c|c}
        \multicolumn{5}{c}{Central to Nearest Neighbor}\tabularnewline
        \hline 
        \hline 
         & Gaussian & Square & CNN-site & CNN-array \tabularnewline
        \hline 
        \hline 
        $\mathcal{F}^{\rm CF}_{52}$ & 0.0298 & 0.0524 & 0.0151 & 0.0099\tabularnewline
        \hline 
        $\mathcal{F}^{\rm CF}_{54}$ & 0.0211 & 0.0298 & 0.0146 & 0.0099\tabularnewline
        \hline 
        $\mathcal{F}^{\rm CF}_{56}$ & 0.0218 & 0.0284 & 0.0120 & 0.0022\tabularnewline
        \hline 
        $\mathcal{F}^{\rm CF}_{58}$ & 0.0392 & 0.0607 & 0.0120 & 0.0020\tabularnewline
        \hline 
        \hline 
        \textbf{$\braket{|\mathcal{F}^{\rm CF}_{5j}|}$} & \textbf{0.0280} & \textbf{0.0428} & \textbf{0.0134} & \bf{0.0060}\tabularnewline
        \hline 
        \multicolumn{5}{c}{}\tabularnewline
        \multicolumn{5}{c}{Between Non-nearest Neighbors}\tabularnewline
        \hline
        \hline
         $\mathcal{F}^{\rm CF}_{13}$ & 0.0110 & 0.0139 & 0.0205 & 0.0110\tabularnewline
        \hline 
        $\mathcal{F}^{\rm CF}_{79}$ & -0.0036 & -0.0005 & -0.0094 & -0.0036\tabularnewline
        \hline 
        $\mathcal{F}^{\rm CF}_{17}$ & 0.0038 & 0.0091 & -0.0016 & 0.0038\tabularnewline
        \hline 
        $\mathcal{F}^{\rm CF}_{39}$ & 0.0028 & 0.0090 & 0.0031 & 0.0028\tabularnewline
        \hline 
         $\mathcal{F}^{\rm CF}_{19}$ & -0.0068& -0.0043 & -0.0098 & -0.0068\tabularnewline
        \hline 
         $\mathcal{F}^{\rm CF}_{37}$ & 0.0070 & 0.0078 & 0.0080 & 0.0070\tabularnewline
        \hline 
        \hline 
        \textbf{$\braket{|\mathcal{F}^{\rm CF}_{ij}|}$} & \bf{0.0058} & \bf{0.0074} & \bf{0.0087} & \bf{0.0058}\tabularnewline
        \hline 
        \hline 
    \end{tabular}
    \caption{Cross-Fidelities at 36 ms readout time in the $5~\mu\rm m$ spacing array. (Top) Cross-fidelity between the central site and the nearest neighboring sites. (Bottom) Cross-fidelity between the edge sites.}
    \label{5um_crosstalk_table}
\end{table} 

At $5~\mu$m separation, the effects of crosstalk become apparent. This is evident if we compare the Square mask cross-fidelities in Table \ref{5um_crosstalk_table} to those in Table \ref{9um-crosstalk-table}. While the edge-to-edge baseline values are similar, the average nearest-neighbour cross-fidelity increased from 0.0138 to 0.0428 when we decreased the spacing from $9~\mu \rm m$ to $5~\mu \rm m$.

\begin{figure*}[!htb]
    \includegraphics[width=\linewidth]{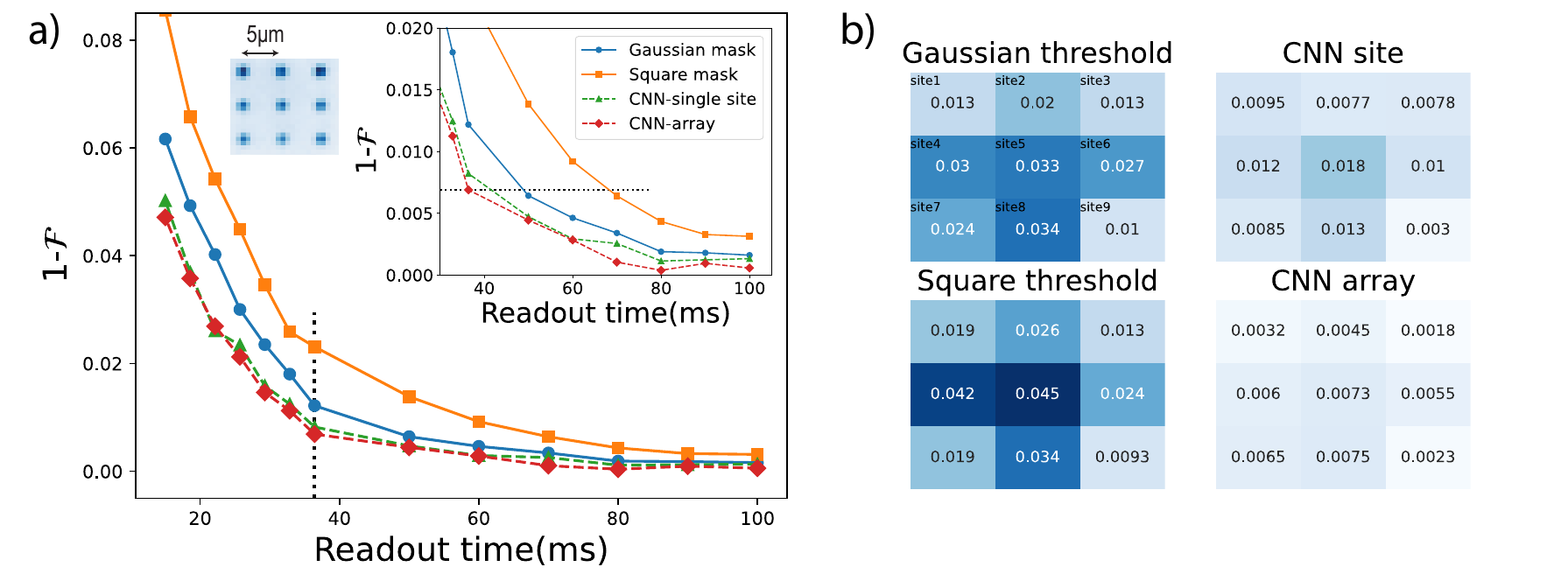}
    \caption{Results from the $5~\mu$m site spacing array. (a) Measurement infidelity with different analysis methods at different readout times. The greatest error reduction is observed at 36 ms, marked with a vertical dashed line. The inset shows the details at readout times above 30 ms. The horizontal dashed line presents the biggest reduction of readout time on CNN-array while maintaining the same fidelity compared to the Gaussian method. (b) Site-per-site infidelity at 36 ms readout time. Higher infidelity is observed around the central site on the array, suggesting the presence of crosstalk from neighboring site fluorescence.  
  }
\label{5um_result}
\end{figure*}

As anticipated, the crosstalk significantly impacts state detection performance, and further differentiates the four methods. Figure \ref{5um_result}a) shows the performance of all four state detection methods for the $5~\mu \rm m$ spaced array. 
CNN-array achieved an infidelity reduction of up to 43\% (vertical dotted line) and readout time reduction up to 25\% (horizontal dotted line) compared to the Gaussian mask. CNN-site achieved a more modest  32\% infidelity reduction and 20\% readout time reduction. 
Unlike the $9~\mu \rm m$ case, the Gaussian mask performs significantly better than the square mask. 

Analyzing the fidelity of individual sites in Fig. \ref{5um_result}b) we observe higher infidelity on the central site when compared to the four corner sites for all methods. The corresponding cross-fidelities in Table \ref{5um_crosstalk_table} support this observation, showing a positive correlation between the central site and its four nearest neighbors. CNN-array was the most effective at mitigating crosstalk, showing almost no difference betwen the nearest-neighbor and edge-to-edge cross-fidelities (0.0060 and 0.0058, respectively). That corresponds to a 78\% cross-fidelity reduction compared to the Gaussian mask. We attribute CNN-array's superior performance to its simultaneous awareness of all sites in the array. Unexpectedly, CNN-site was also successful at mitigating some of the crosstalk, achieving a 50\% cross-fidelity improvement. We attribute this to CNN-site's spatially-aware utilization of features on the edges of the $10\times 10$ pixel input to detect subtle correlations.

\begin{figure}[!htb]
    \centering
    \includegraphics[width=\linewidth]{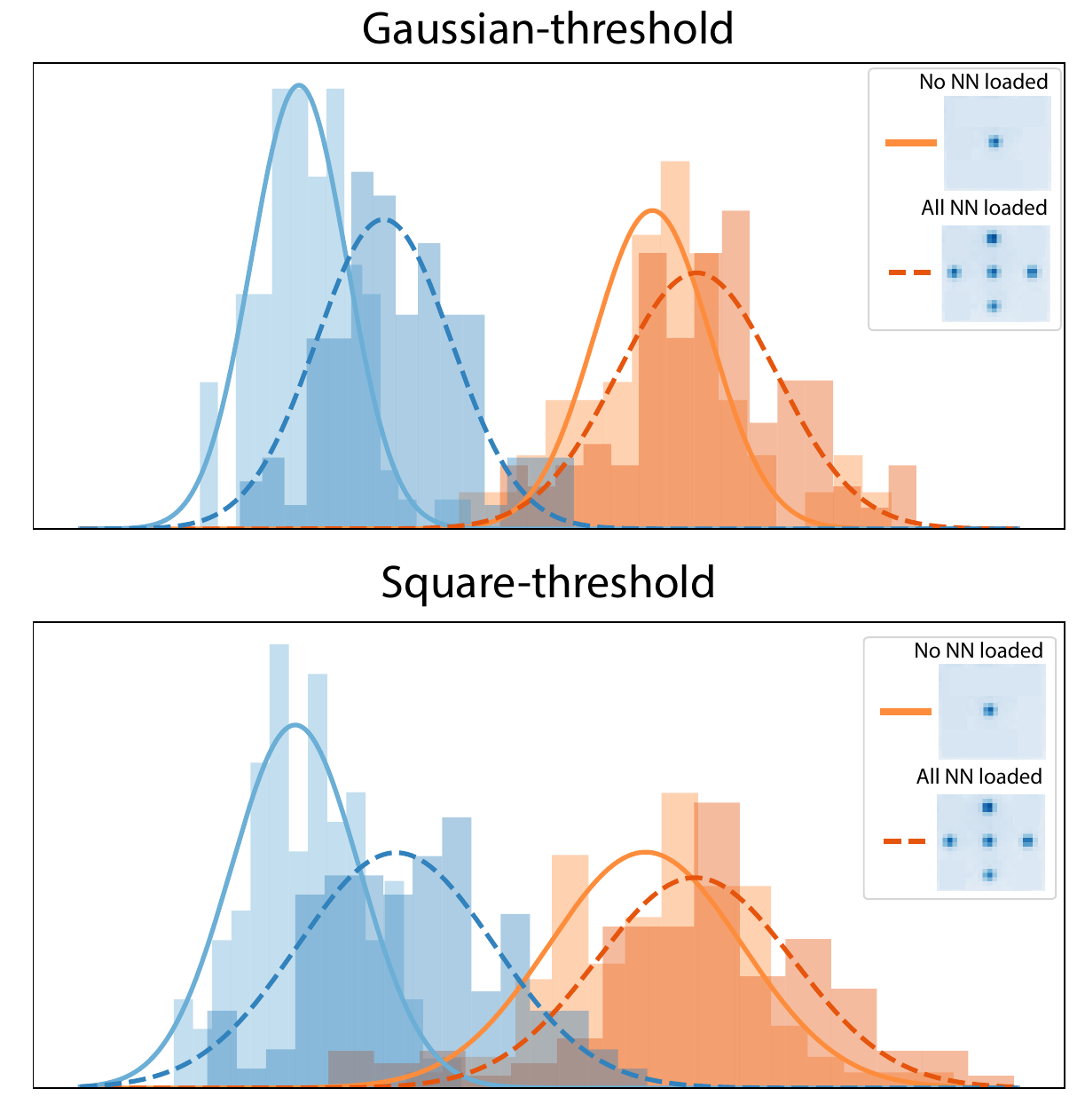}
    \caption{The shift of the integrated distribution due to the absence and presence of atoms in neighboring sites on the Gaussian mask(top) and Square mask(bottom). The overlap between the two states substantially increased in the square mask method, resulting in poor fidelity.}
    \label{5um_crosstalk_histogram}
\end{figure}

To better understand the effect of crosstalk on state detection performance, we compare in Fig. \ref{5um_crosstalk_histogram} the histograms of the central site with and without neighbors for the Gaussian and Square threshold methods. When neighboring sites were occupied, we observed broadening and upward-biasing of the distributions due to bleed-over of fluorescence from neighboring sites, making the two states more difficult to separate. 

\rsub{Finally, in order to gauge the advantages of our method under normal operating conditions, we apply the CNN detection method to measurements taken on the non-attenuated primary path at $5~\mu \rm m$ spacing. We obtain the training data and ``ground truth'' labels by performing a sequence of three measurements on each randomly loaded atom array, as described in Appendix \ref{s_singlepath_dataset}. The first and last high fidelity measurements are 20 ms long with $>4.5\sigma$ separation. We observed improvement with up to 83\% reduction of infidelity. This improvement is consistent with results from the dual-path setup. This makes it possible to reduce the probing time from 15 ms to 9.8 ms while preserving the readout fidelity above 99.5\%.} 

\rsub{We also characterized the  processing time needed by the CNN compared to the conventional method. On a PC (Ryzen 1700 CPU and GTX1080 GPU), we found the average CNN-site and CNN-array single-site inference time is 97 $\pm$ 1 $\mu \rm s$ and 303 $\pm$ 23 $\mu \rm s$ respectively, compared to the conventional processing time of 11 $\pm$ 1 $\mu \rm s$. Taking into  account the increased inference time together with the reduced probing time, our method reduces the total measurement time by 50\%. The processing delay can be further reduced by optimizing the CNN network and using hardware acceleration.}

\begin{figure}[!htb]
    \centering
    \includegraphics[width=\linewidth]{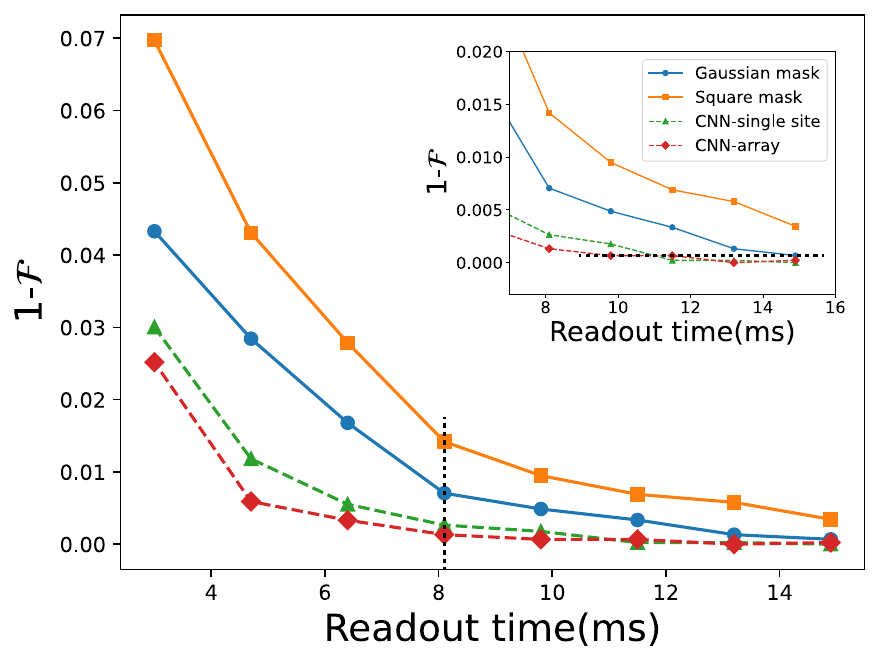}
    \caption{\rsub{Measurement results from the primary non-attenuated path for a $3 \times 3$ array with 5 $\mu \rm m$ spacing. The greatest fractional error reduction between CNN-site and the Gaussian method is marked with a vertical dashed line. The inset shows the details for readout times above 7 ms. The horizontal dashed line shows that CNN-site at 11.5 ms  achieves the same fidelity as the Gaussian method at 15. ms.}}
    \label{single_path_result}
\end{figure}

\section{Conclusion}
We have presented two neural network architectures for state detection of neutral atoms. Both architectures outperformed conventional state detection methods by a statistically significant margin. We observed a 43\% reduction in infidelity and 78\% cross-fidelity reduction at $5~\mu\rm m$ site-to-site spacing with CNN-array, demonstrating a superior ability to mitigate inter-site crosstalk. CNN-site also performed better than conventional methods, with a 32\% reduction in infidelity and 50\% cross-fidelity improvement. To confirm that improvement is not attributable only to the mitigation of crosstalk, we also evaluated the performance of the neural network at 9 $\mu$m site spacing where the effects of crosstalk are small. In this configuration, CNN-site was able to reduce the infidelity up to 57\% compared to the Gaussian mask method.  Based on these results, we conclude that a neural network can reduce the readout time by up to 29\% while maintaining the same fidelity, opening the possibility of faster high-fidelity measurements on neutral atom processors. In future work we plan to accelerate the CNNs on an FPGA as a proof-of-concept demonstration of fast real-time state detection, which is a necessary ingredient for measurement-based quantum error correction\cite{Graham2023b}.

\acknowledgments
This material is based upon work supported by the U.S.
Department of Energy Office of Science National Quantum
Information Science Research Centers as part of the Q-NEXT
center, NSF Award 2016136 for the QLCI center Hybrid Quantum Architectures and Networks, and NSF award No. 2210437.

\bibliography{atomic,saffman_refs,rydberg,qc_refs,optics}

\appendix 

\section{Methods} \label{S_training}
This section describes the step-by-step procedure used to initialize and train the neural networks.

\subsection{Preprocessing} \label{s_preprocess}
The primary- and secondary-path images are acquired from the camera as raw, greyscale images. The primary-path images are fed into the conventional Gaussian mask analysis method (see Sec. \ref{conventional_methods}) in order to extract the ``ground truth'' labels -- a binary value for each site in the array, where `1' corresponds to the bright state and `0' to the dark state. This results in a binary array of size $(n_{\rm images}$, $n_{\rm sites})$, encompassing all the labels for the entire dataset. Prior to being sent to the CNN for processing, the secondary-path images are first normalized and zero centered: $I_{ij}$=($I_{ij}-\mu)/\alpha$ where $\mu$ is the average pixel intensity over the entire training subset and $\alpha=I_{\rm max}-I_{\rm min}$ where $I_{\rm max}$ and $I_{\rm min}$ are maximum and minimum pixel intensity in the training subset. The indices $i$ and $j$ refer to the coordinates of each pixel in the image.

\begin{figure}[!t]
    \includegraphics[width=\columnwidth]{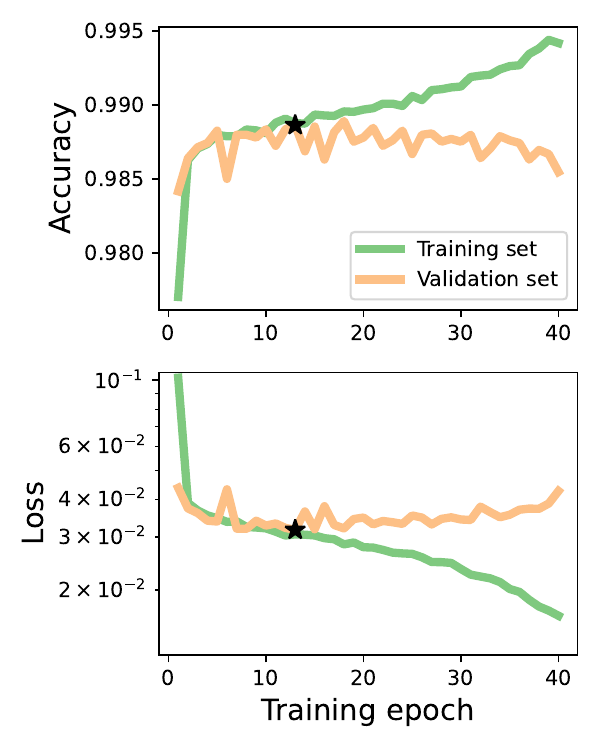}
   \caption{The accuracy and loss over the training epoch. The optimum parameter set is extracted from the epoch with the lowest validation loss, marked in the figure as $\star$. }
    \label{s_training_history}
\end{figure}

\subsection{Image Partitioning (only for CNN-site)}
The input images for CNN-site are generated by cropping the array image into individual $10\times10$ pixel images at predefined site locations. For each site, the site center is determined by performing a Gaussian fit over the averaged training image. Averaging the training images improves the signal-to-noise ratio and ensures that all sites contain the (averaged) image of an atom. Once the pixel coordinates at a site's center have been identified, a $10\times10$ crop centered on those coordinates is extracted. This is repeated for all sites in the array. All of the extracted site images together form an array of shape $(n_{\rm image}$, $n_{\rm site}$, 10, 10), where $n_{\rm image}$ refers to the number of original (un-cropped) images. This array is reshaped to a 4-D matrix of size $(n_{\rm image}\times n_{\rm site}$, 10, 10, 1), which is passed to the CNN as input data (Tensorflow processes the images in batches, not one at a time). A separate array of this type is produced for each of the training, test, and validation subsets. The array of ``ground truth'' labels is transformed to a 1-D vector of size ($n_{\rm image}\times n_{\rm site}$). The CNN does not receive these labels as input; they are used by Tensorflow to evaluate the CNN and manage the training process.

\subsection{Training}
The network is trained using the Tensorflow library, via the Keras API. The network parameters are randomly initialized to begin with, then trained using the pre-processed secondary-path images and corresponding ``ground truth'' labels. The hyper-parameters used for training are listed in Table \ref{parameter_table}. The training progress is supervised by monitoring the loss over the validation set at the end of every training epoch, and saving a snapshot of the best-so-far network parameters. At the end of the training process, only the set of parameters corresponding to the lowest-achieved validation loss is kept. Validation loss refers to the loss metric evaluated on the validation subset. A lower validation loss corresponds to better performance at the task. The max epoch number is chosen to be long enough to avoid underfitting the data. In other words, the training continues until overfitting occurs, such that the training accuracy will approach 100\% by the end of the training process, while the validation accuracy begins to drop. Training to the point of overfitting does not harm the final result, because of this monitoring process. Roughly speaking, the optimal network parameters are achieved when the training accuracy (loss) and validation accuracy (loss) diverge, as seen in Fig. \ref{s_training_history}. This process of monitoring and selection ensures that  the network parameters are a sufficiently good fit to the classification task and that the resulting network still generalizes well to data that it has not seen before.

 \begin{table}[!ht]
    \centering
    \begin{tabular}{|c|c|c|} 
        \hline 
        Parameter & CNN-site & CNN-array  \tabularnewline
        \hline 
        \hline 
        Optimizer & Adam & Adam\tabularnewline
        \hline 
                Initial learning rate & $1\times10^{-4}$ & $5\times10^{-4}$ \tabularnewline
        \hline 
        Max epoch & 40 & 30\tabularnewline
        \hline 
        Batch size & 64 & 16\tabularnewline
        \hline 
    \end{tabular}
    \caption{Training hyperparameters. 
    Adam\cite{Kingma2017} is a stochastic gradient descent algorithm with adaptive learning rates from estimation of the first and second order of the gradients. Initial learning rate determines the step size of the first parameter updates towards the minimum of a loss function. The learning rate for the further iterations are dynamically determined by the Adam optimizer. Max epoch is the number of complete passes through the training dataset during training. The batch size is the number of  samples passed before the network parameters are updated during training.  }
    \label{parameter_table}
  \end{table}

Note that the CNN needs to be re-trained from scratch for every new dataset acquired. For example, CNN-site has a completely different set of training parameters for the ``$9~\mu \rm m$, 40 ms readout time'' dataset as compared to the ``$9~\mu \rm m$, 20 ms readout time'' dataset or the ``$5~\mu \rm m$, 20 ms readout time'' dataset. We found that it is not difficult to modify the CNN architecture in such a way (namely, by incorporating an ``embedding'' layer and providing a dataset identifier as an additional input) that a single parameter set can handle all datasets equally, without significant loss of accuracy. This could be an important enhancement for real-time applications in which it is desirable to change the parameters of the experiment without acquiring a new dataset and re-training the CNN. We consider this to be a topic for future research.

\section{\rsub{Alternative dataset generation method}} \label{s_singlepath_dataset}
\rsub{For a system with only a single imaging path, a training dataset can be generated by performing a sequence of $A$-$B$-$A$ measurements on each randomly loaded array, where $A$ is a high fidelity measurement and $B$ is a noisy measurement. Then the data are post-selected to keep only the $A$-$B$-$A$ sequences in which the first and last $A$ measurements show identical array occupation (no atom loss), according to the conventional analysis method. Post-selection discards any data that are mislabeled due to atom loss during or after the first measurement. The post-selected $A$ measurements are then used to generate the ``ground truth'' labels, while the corresponding $B$ measurements constitute the inputs to the CNN. }

\end{document}